\documentclass[twocolumn,aps,prd,nobibnotes,byrevtex,nofootinbib,superscriptaddress,amsfonts,amsmath,amssymb,floatfix,10pts]{revtex4-1}
\usepackage{hyperref}
\usepackage{tablefootnote}
\usepackage{enumerate}
\hypersetup{
    colorlinks=true,
    citecolor=blue,
    linkcolor=blue,
    filecolor=magenta,      
    urlcolor=blue,
}
\usepackage{amsmath}
\usepackage{charter}
\usepackage{subfig}
\usepackage{graphicx}
\usepackage{multirow}
\usepackage{dcolumn}
\usepackage{mathrsfs}
\usepackage{acronym}
\usepackage[usenames,dvipsnames,svgnames]{xcolor}
\usepackage{soul}
\usepackage{tabularx}

\begin{document}
\title{Identifying multichannel coherent couplings and causal relationships
in gravitational wave detectors}
\author{Piljong Jung}
\affiliation{Gravity Research and Application Team (GReAT), National Institute for Mathematical Sciences (NIMS), Republic of Korea}

\author{Sang Hoon Oh}
\affiliation{Gravity Research and Application Team (GReAT), National Institute for Mathematical Sciences (NIMS), Republic of Korea}

\author{Young-Min Kim}
\affiliation{Department of Physics, Ulsan National Institute of Science and Technology (UNIST), Republic of Korea}

\author{Edwin J. Son}
\affiliation{Gravity Research and Application Team (GReAT), National Institute for Mathematical Sciences (NIMS), Republic of Korea}

\author{Takaaki Yokozawa}
\affiliation{Institute for Cosmic Ray Research (ICRR), KAGRA Observatory, The University of Tokyo, Japan}

\author{Tatsuki Washimi}
\affiliation{Gravitational Wave Science Project (GWSP), Kamioka branch, National Astronomical Observatory of Japan (NAOJ), Japan}

\author{John J. Oh}
\email[]{johnoh@nims.re.kr}
\affiliation{Gravity Research and Application Team (GReAT), National Institute for Mathematical Sciences (NIMS), Republic of Korea}

\date{\today}


\begin{abstract}
The gravitational-wave detector is a complex and sensitive collection of advanced instruments that are impacted not only by mechanical/electronics systems but also by the surrounding environment. Hence, it is of great importance to classify and mitigate noises to detect gravitational-wave signals by using information from many auxiliary channels related to such devices and surroundings. This improves the signal-to-noise ratio and reduces false alarms from coincident loud events. For this reason, it is essential for identifying coherent relationships between complex channels. This study presents a way of identifying (non-) linear couplings between associated channels by using the method of correlation coefficients. And we show that the method can be applied to practical problems in the gravitational-wave detector, such as noises by lightning strokes, air compressors vibrations, and noises caused by wind effects.
\end{abstract}

\maketitle


\acrodef{gw}[GW]{gravitational wave}
\acrodef{ligo}[LIGO]{Laser Interferometer Gravitational-wave Observatory}
\acrodef{snr}[SNR]{signal-to-noise ratio}
\acrodef{pcc}[PCC]{Pearson's correlation coefficient}
\acrodef{kendall}[Ktau]{Kendall's $\tau$ coefficient}
\acrodef{mic}[MIC]{Maximal Information Coefficient}
\acrodef{mi}[MI]{mutual information}
\acresetall

\section{Introduction}
\label{sec:intro}
The detection of \ac{gw} emitted from the binary black hole coalescence (GW150914) \cite{Abbott:2016blz} opened the era of \ac{gw} astronomy, which is expected to pursue new perspectives of understanding the structure of the Universe and the evolution of astrophysical objects such as stars and galaxies. For a more profound understanding and expansion of the knowledge of the Universe, there still exist many challenges to be overcome for \ac{gw} physics and astronomy. Much more accurate sky localization for \ac{gw} sources and a much farther range of luminosity distance in \ac{gw} observation requires the improved sensitivity of \ac{gw} detectors, overcoming the present limitation of technologies.

Detectors' sensitivity confines the boundary of observations so that the number of \ac{gw} events within the boundary determines the detection rate that can be achieved in the current \ac{gw} detector. The currently operating ground-based laser interferometric \ac{gw} detectors such as \ac{ligo}, Virgo, and the KAGRA have a similar design sensitivity curve with a frequency band of $30-2000Hz$, which is characterized by three primary noise sources: photon shot noise by the laser system in the high-frequency range, thermal noise by test-mass mirrors in the mid-frequency range, and seismic noise by ground vibrations in the low-frequency range \cite{TheLIGOScientific:2014jea, TheVirgo:2014hva, Akutsu:2018axf, KAGRA:2013pob}. Those \ac{gw} detectors are now being operated for detecting \ac{gw} signals and planned for upgrades to improve their sensitivities using many engineering challenges such as cryogenics, quantum squeezed light, and so on. Furthermore, next-generation ground-based \ac{gw} detectors with new conceptual designs are now being planned \cite{Punturo:2010zz, Evans:2016mbw}.

Besides, advanced methodologies and analysis algorithms should be required to enhance the quality of data \cite{Aasi:2014mqd, TheLIGOScientific:2016zmo, Jung:2018adn} taken from the very sensitive instruments and isolate \ac{gw} signals from noises caused by the sensitive devices and the surrounding environments efficiently because it improves the detection statistics such as \ac{snr} to provide the reliable detection criterion. This, consequently, also improves the detection range of \ac{gw} detectors, which yields better detection rates of \ac{gw} sources. In this point of view, both developments of advanced methodologies for data analysis and understanding the status of \ac{gw} detectors, as well as the most sensitive instruments, are of special importance to achieve for detecting \ac{gw} signals coming from the farthest \ac{gw} emitting objects.
For this reason, many different tools for characterizing noises have been developed and utilized for analyzing \ac{gw} signals so far. The first purpose of those tools is to categorize and classify transient/continuous noises that are harmful to the \ac{gw} strain channel, identifying the causal relationship in the coincident families of noisy channels. Then they should be mitigated with the help of various advanced mathematical algorithms as possible. Finally, if they are caused by some instrumental defects with a possibility of repeated malfunctions, the causes should be reported and amended for maintaining the consistent status of the detector.

Studies of finding coherence between the strain channel and auxiliary channels of the \ac{gw} detectors have been extensively performed in \ac{ligo}-Virgo collaborations, in the context of continuous \ac{gw} and stochastic \ac{gw} background searches \cite{LSC:2018vzm, LIGO:2021ppb, Christensen:2010zz, LIGOScientific:2014qfs, VIRGO:2012oxz, Acernese_2007}.
Many efforts for identifying and vetoing transient noises have been made so far and are widely utilized for \ac{gw} data analysis, such as a computation of significance called {\it hierarchical veto (Hveto)} \cite{Nuttall:2015dqa} and {\it Used percentage veto (UPV)} \cite{Isogai:2010zz}, noises associated with long-duration transients \cite{Prestegard:2011me, PhysRevD.83.083004}, Q-transform based trigger generator \cite{Robinet:2020lbf}, Hilbert-Huang transform-based method \cite{Son:2018xpr}, bicoherence method \cite{Jaranowski:2005hz}, linear regressions \cite{Walker:2018ylg}, machine learning algorithms \cite{Cuoco:2020ogp}, and so on. These methods cover classification and a vetoing method as well as identification of the influences between the \ac{gw} strain channel and auxiliary channels monitoring the environmental and/or instrumental status. This needs a consistent understanding of which channel effects can cause the transient noises in the \ac{gw} strain channel and/or which can be useful to mitigate such abnormalities to maintain the normal status of the detector. However, we still face great challenges in dealing with the noises of \ac{gw} data because the \ac{gw} detector behaves with a highly non-stationary and non-linear nature. 

For this reason, it is necessary to develop more advanced analysis tools for noise hunting to improve the detector's data quality. Here we focus on the couplings between a certain auxiliary channel and the \ac{gw} strain channel. In Ref. \cite{Bose:2016sqv}, it has been shown that the excess noises from bilinear and non-linear couplings in \ac{gw} interferometers can be treated using the bilinear coupling veto (BCV) method \cite{Ajith:2014aea} with the trigger-based correlation coefficient.
Along this line, we suggest a new way of identifying coherent associations between the \ac{gw} strain channel and the related auxiliary channels, in which we use three kinds of the correlation coefficient: \ac{pcc} \cite{Pearson} and \ac{kendall} \cite{Kendall} as a linear measure and \ac{mic} as a non-linear measure \cite{reshef_detecting_2011, mice2016, mice2018}, respectively. In particular, \ac{mic} is an information-theoretic measure to discriminate the non-linear association between two random variables. Together with these measures, we construct a systematic way of identifying the noises from (non-)linear couplings and causalities propagating from instrumental and/or environmental disturbances of \ac{gw} detectors. Then we apply the suggested method to the well-known issues of data analysis and noise identification in the KAGRA detector, expecting that the correlation unidentified in the previous methods can be verified as a non-linear correlation effect.

In this study, we present a way of identifying coherent associations between the \ac{gw} strain channel and auxiliary noise channels by computing correlation measures. The method we consider here is the \ac{pcc} and \ac{kendall} as a linear measure and \ac{mic} as a non-linear measure. We construct a consistent way of discriminating the relevant noise effect and its causality, applying it to some issues in \ac{gw} detection. In Section \ref{sec:cagmon}, we describe the methods to measure (non-)linear associations and build an analysis process with proper statistical algorithms between the \ac{gw} strain channel and many auxiliary channels. In Section \ref{sec:nldetection}, we exhibit the exemplary results based on some noise data taken from the KAGRA \ac{gw} detector; a lightning stroke, air compressor noises, and the noises caused by the wind effect. Finally, we summarize and discuss our results in Section \ref{sec:discussion}.
 
\section{Method and Workflow}
\label{sec:cagmon}
We use the data taken from the \ac{gw} strain channel and many auxiliary channels in the KAGRA \ac{gw} detector. The KAGRA is a gravitational-wave detector with a similar configuration of laser interferometry such as the \ac{ligo}/Virgo except for the cryogenic test-mass mirrors and the underground installation \cite{10.1093/ptep/ptaa125}. The initial installation of the KAGRA has finished in 2019, and after year-long commissioning, it started its first observing run during a month in 2020 and joined the O3 observing run together with the advanced \ac{ligo} and the Advanced Virgo, recently \cite{Kokeyama:2020dkg}. 

The major feature of the KAGRA is the cryogenic and underground \ac{gw} detector, which implies that the KAGRA detector has somewhat unique characteristics caused by the nature of the underground cryogenic facility, producing the relevant noise effects that were not reported in the ground-based \ac{gw} detectors. Therefore, it is of great importance to understand the noise characteristic of the KAGRA detector and its environment.

\subsection{Methods for Data Correlations}
We introduce three methods of analysis for investigating the correlation between two data samples; the \ac{pcc}, the \ac{kendall}, and the \ac{mic}.
Let us consider two time-series data $X$ and $Y$ to be non-stationary and uni-variate data sets with an equal size of $n$. We assume that $(x_i,y_i)$ is a set of $i$-th bi-variate data pairs from the paired data $(X,Y)$. If we assume that the data set $Y$ includes certain noises, being affected by the sensitivity of \ac{gw} detectors, there exists a meaningful statistical association between two observed variables $X$ and $Y$ because the noise can propagate to another data set $X$.
With these assumptions, we analyze the correlated relationship of noises resulting from the instrumental anomalies and/or environmental interference in \ac{gw} detectors. In this section, we describe three major methods to estimate the linear and non-linear associations based on the time-series data recorded in each channel.

The \ac{pcc} is defined as the ratio between the covariance and the product of standard deviation of each variable, which produces the linear correlation score. More precisely, let $\bar{x}=\sum_{i=1}^{n}x_i/n$ and $\bar{y}=\sum_{i=1}^{n}y_{i}/n$ be the means of $X$ and $Y$, respectively, then, the PCC $\rho$ is 
\begin{equation}
\rho(X,Y)=\frac{{\sum_{i=1}^{n} (x_i - \bar{x})(y_i - \bar{y})}}{{\sqrt{\sum_{i=1}^{n} (x_i - \bar{x})^2 \sum_{i=1}^{n} (y_i - \bar{y})^2}}},
\end{equation}
which determines the value between $-1$ and $1$. For independent two variables, we have $|\rho|=0$. When $|\rho| > 0.5$, it is typically interpreted as a significant correlation. $|\rho|=1$ indicates that two variables have a perfect linear correlation, where positive and negative signs imply increasing and decreasing linear dependence, respectively. 
Here, we used the absolute value of the \ac{pcc} for convenience.

The \ac{kendall} \cite{Kendall} measures the strength of the monotonicity of the relationship between two variables, which is defined by
\begin{equation}
\tau(X,Y) = \frac{2(c-d)}{n(n-1)}, 
\end{equation}
where $c$ and $d$ are the number of concordant and discordant pairs in $(X,Y)$, respectively. Given two data samples from the combined variable set: $(X,Y)=\left\{(x_1, y_1), (x_2,y_2), \cdots, (x_n, y_n) \right\}$, if sampled pairs are  either $x_i > x_j$ and $y_i > y_j$ or $x_i < x_j$ and $y_i < y_j$, it is called a concordant pair. On the other hand, if they are either  $x_i > x_j$ and $y_i < y_j$ or $x_i < x_j$ and $y_i > y_j$, it is called a disconcordant pair. Hence, \ac{kendall} provides the ordinal association that is proportional to the difference between ordered and disordered pairs in all possible combinations.
As with the PCC, $\tau(X, Y)$ varies from $-1$ to $1$. If the order of two pairs is randomly distributed, they are monotonically independent, yielding $\tau(X, Y)=0$. If $Y$ values tend to change with increasing or decreasing $X$ values, the absolute value of \ac{kendall} becomes one. 

Meanwhile, the \ac{mi} can estimate and characterize the strength of shared information between two random variables. For given two discrete random variables $A$ and $B$ with a joint probability mass function $p(a,b)$ and marginal probability mass functions $p(a)$ and $p(b)$, \ac{mi} is defined by
\begin{equation}
I(A;B) = \sum_{a \in A}\sum_{b \in B}p(a,b) \log_2{\frac{p(a,b)}{p(a)p(b)}}.
\end{equation}
Note that \ac{mi} provides non-negative values, $I(A;B) \ge 0$. If the random variable $B$ is a function of $A$, $B=f(A)$, $I(A;B)$ diverges. In addition, if $A$ and $B$ have no shared information, $p(a,b)=p(a)p(b)$, then $I(A;B)$ clearly vanishes; they are statistically independent. Refer \cite{mitextbook} for more detailed properties of \ac{mi}.

Suppose that the joint probability distribution $p_{XY}(i, j)$ is approximated by the number of points falling into the $i$-by-$j$th bin on the partitioned plane of scattered plots by $X$ and $Y$. Then, the approximated \ac{mi} is obtained by the occupied elements in each cell as
\begin{equation}
    \label{eq:mi}
I(X;Y)_{k,l} = \sum_{i=1}^{k}\sum_{j=1}^{l}\frac{p_{XY}(i,j)  }{ \log_2{\min \left\{ k, l \right\}} }\log_2{\frac{p_{XY}(i,j)}{p_{X}(i)p_{Y}(j)}},
\end{equation}
where $p_{X}(i)$ and $p_{Y}(j)$ are marginal distributions on $i$-th column and $j$-th row. In addition, $k$ and $l$ indicate the partition size of $X$ column and $Y$ row, respectively. Because the Jensen's inequality, $0 \le I(X;Y)_{k,l} \le  \log_2{\min \left\{ k, l \right\}}$, is satisfied \cite{mitextbook}, Eq. (\ref{eq:mi}) can be normalized and the value spans between zero and one.

To overcome a heuristic approach of maximizing \ac{mi} value for all possible resolutions of $k$-by-$l$ grid \cite{reshef_detecting_2011}, three intrinsic properties in Eq. (\ref{eq:mi}) are utilized; they are 1) monotonic convex function, 2) bounded set, and 3) uniformly continuous function, which allows computing Eq. (\ref{eq:mi}) more efficiently and effectively.

\begin{figure}[t!]  
\begin{center}
\includegraphics[width=\columnwidth]{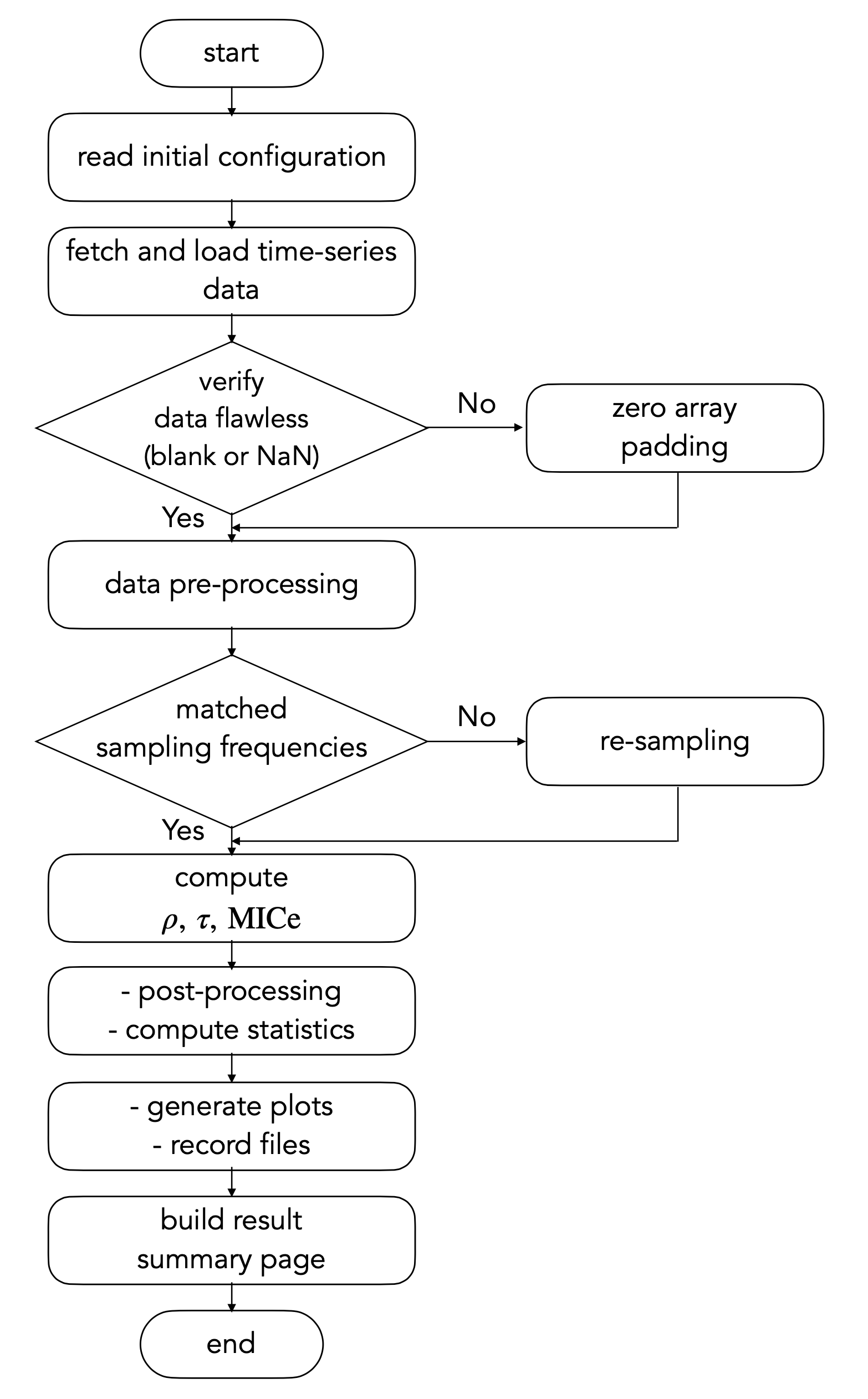}  
\caption{\small A flowchart of the {\it CAGMon} tool.} \label{Fig.algorithm}
\end{center}
\end{figure}

\begin{figure*}[t!]  
\begin{center}
\includegraphics[width=\textwidth]{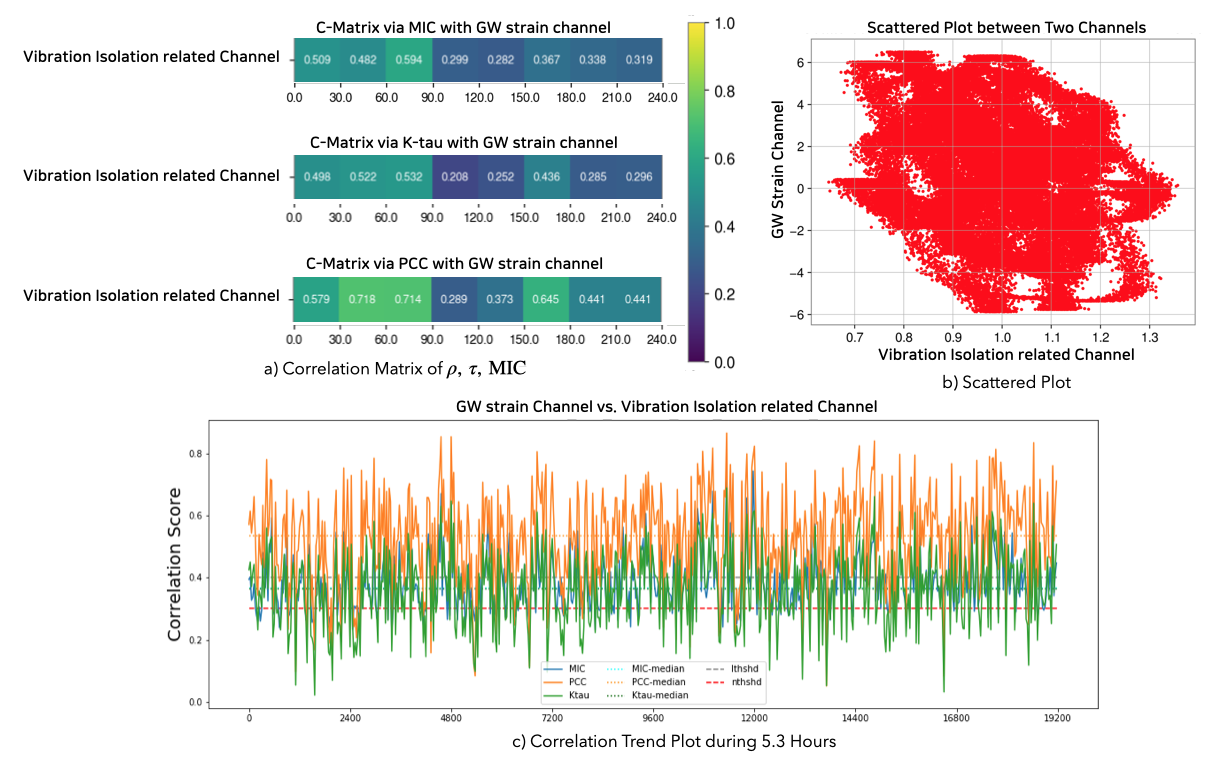}  
\caption{\small Exemplary plots provided by the {\it CAGMon} tool: a) correlation matrix plot of $\rho,~\tau$, and ${\rm MIC}$ between \ac{gw} strain channel and auxiliary channels b) scattered plot of both channels (Each relative amplitude is rescaled by its median.) c) correlation trend plot within a certain duration. Here, we demonstrate one of vibration isolation system (VIS) channels among tens of auxiliary channels in the KAGRA \ac{gw} detector as a simple example.} \label{Fig.cagmon}
\end{center}
\end{figure*}

Let $S$ be a $n$-ordered pair set and $B(n)=n^{\alpha}$ and the parameter $c$ restricts the total grid size. Here, $\alpha$ is a dimensionless parameter that controls the size of grids and $c$ is a controlling the coarseness of maximizing the discrete grid search. Then, the empirical \ac{mic} called MICe is defined as
\begin{equation} \label{eq:mice}
MICe(X, Y, \alpha, c) = \max_{ab<B(n)} \Big\{\frac{\max I^{[*]}(S, k, l)}{\log_{2} \min\{k,l\}} \Big\},
\end{equation}
where $I^{[*]}(S, k, l)$ denotes the maximized \ac{mi} in $k \times l$ grids.
Note that the coefficient of MICe varies from zero to one. If two time-series data have a clear association, Eq. (\ref{eq:mice}) converges to one for sufficiently large data sizes whereas it becomes to zero in the opposite condition. Even though MICe has a unique performance to extract the complex associations, it was uncertain how reliable MICe values can be provided in the aspect of the parameter selection. To guarantee the reliability of computed MICe, we suggested an empirical way of determining the relevant parameters of MICe through the statistical power to optimize them in \cite{Jung:2021jss}.


\subsection{CAGMon: A Novel Tool for Identifying Correlations} \label{subsec.cagmon}

We utilize the aforementioned correlation measure indices to identify the data associations between the \ac{gw} strain channel and auxiliary channels of GW detectors. To monitor the chronological trend of association coefficients between two datasets, we considered a minimal bin of data segments, called {\it stride}, which represents a piece of an equal interval unit in the whole data segments. For given data sets, the aforementioned three correlation coefficients, $\left\{ \rho(X_m, Y_m), \tau(X_m, Y_m), MICe(X_m, Y_m) \right\} $, are computed for divided subsets of $X_m$ and $Y_m$ by stride. By gathering the sequential coefficients together, we can demonstrate a time-serial trend of the association strength between two channels. In this section, we present a novel tool for identifying and diagnosing the association of two datasets, called {\it CAGMon}; the workflow and design of the algorithm.

The workflow of {\it CAGMon} tool is depicted in Fig. \ref{Fig.algorithm}, which comprises four different stages: i) reading initial configuration, ii) loading time-series data and pre-processing iii) computing each coefficient and the relevant statistics, and iv) plotting results and building a result summary page. 
First, the configuration file consists of user-defined parameters, pre-processing options, and general features of data such as sampling rate, start/end times, and stride. Because each coefficient is computed for each stride of time-series datasets, the tool can only investigate the similar timing coincidence within the same stride bin. Hence, there has a limitation in identifying the association between transient noises correlated to noise with a specific frequency. To overcome this limitation, we provide several customized options for data pre-processing, such as high-pass, low-pass, and bandpass filters, for a variety of scalable analyses. In addition, the general feature of data comprises a primary channel name, the path of data files, an auxiliary channel list file, segment file, output file type and its save path, and so on.
When the time-series data is retrieved from \ac{gw} frame files, the configuration option refers to the initial configuration file to apply the pre-processing. If time-series data contain either blank or NaN, zero arrays would be padded properly to avoid a computational error. Then, data are re-scaled to have the same data size using the resampling algorithm. 
The detailed discussion and analyses on the reliability of MICe values, data sample size, and optimal parameters of MICe have been presented in \cite{Jung:2021jss}. In this study, we select a set of optimal parameters of MICe presented in Table \ref{tab:parameters}.

In Fig. \ref{Fig.cagmon}, the {\it CAGMon} tool presents: i) a correlation matrix of $\rho$, $\tau$, and \ac{mic} between two datasets, ii) a scattered pattern between both channels, and iii) the correlation trend behavior during a certain time period between two channels. The correlation method simply produces the coherence between two datasets. So one requires a way of improving the statistical significance of a detected observation. Quantitatively, the way of using timeslide studies in \cite{Was:2009vh, Kowalska-Leszczynska:2016low} can implement the statistical significance of the coincident events. Originally, the method is based on the signal-to-noise ratio (SNR) of the matched filter, but in our case, we could incarnate it with the correlation score instead of using the SNR. Then we could conduct the timeslide by shifting a stride back and forth for a given high-scored event, counting the number of events.

Qualitatively, an alternative way is to refer to the subsystems that affect each other. When we investigate a family of correlated subsystems that is closely related, because some similar observations can be seen from the different related channels, one can estimate the observed correlation with a reliable statistical significance, at least, qualitatively.
More precisely, when we investigate the effect of magnetic fields, the magnetometer-related channels with a family of the same subsystem (MAG-subsystem) are affected with similar behaviors, i.e., a similar correlation appears. Therefore, we could conclude that the correlation index in the magnetometer-related channels reveals a piece of reliable evidence with the statistical significance of the observation. A specific example can be found in the applications in the following section.

\renewcommand\arraystretch{1.6}
\renewcommand{\tabcolsep}{0.8pt}
\begin{table}[t!]
\begin{tabular}{cccccc}
\hline \hline
channel type & stride(s) & sampling rate(Hz) & data size & $\alpha$ & $c$  \\ \hline\hline
\ac{gw} strain  & 2 & 4096 & 8192 & 0.5 & 0.7 \\ 
BNS range & 512 & 16 & 8192 & 0.5 & 7.0 \\ \hline
\end{tabular}
\caption{\small The channel information and parameters of MICe used in this study: these values are selected by the empirical optimized parameter selection in \cite{Jung:2021jss}.}
\label{tab:parameters}
\end{table}

\section{Application to Gravitational-Wave Data} 
\label{sec:nldetection}
{\it CAGMon} tool was designed to identify and diagnose associated auxiliary channels that influence the \ac{gw} strain channel for \ac{gw} detection, producing a time-sequential correlation trend propagated by instrumental or environmental disturbances. We can estimate a significant association as excess from the long duration trend of its median value. 
In this section, we apply the {\it CAGMon} tool to the KAGRA \ac{gw} data for characterizing and identifying associations between channels caused by well-known environmental and instrumental events such as lightning stroke and air compressor noises.

\subsection{Magnetic Field Transients from Lightning stroke}  
\label{sec:lightning}
\begin{figure}[t!]  
\begin{center}
\includegraphics[width=\columnwidth]{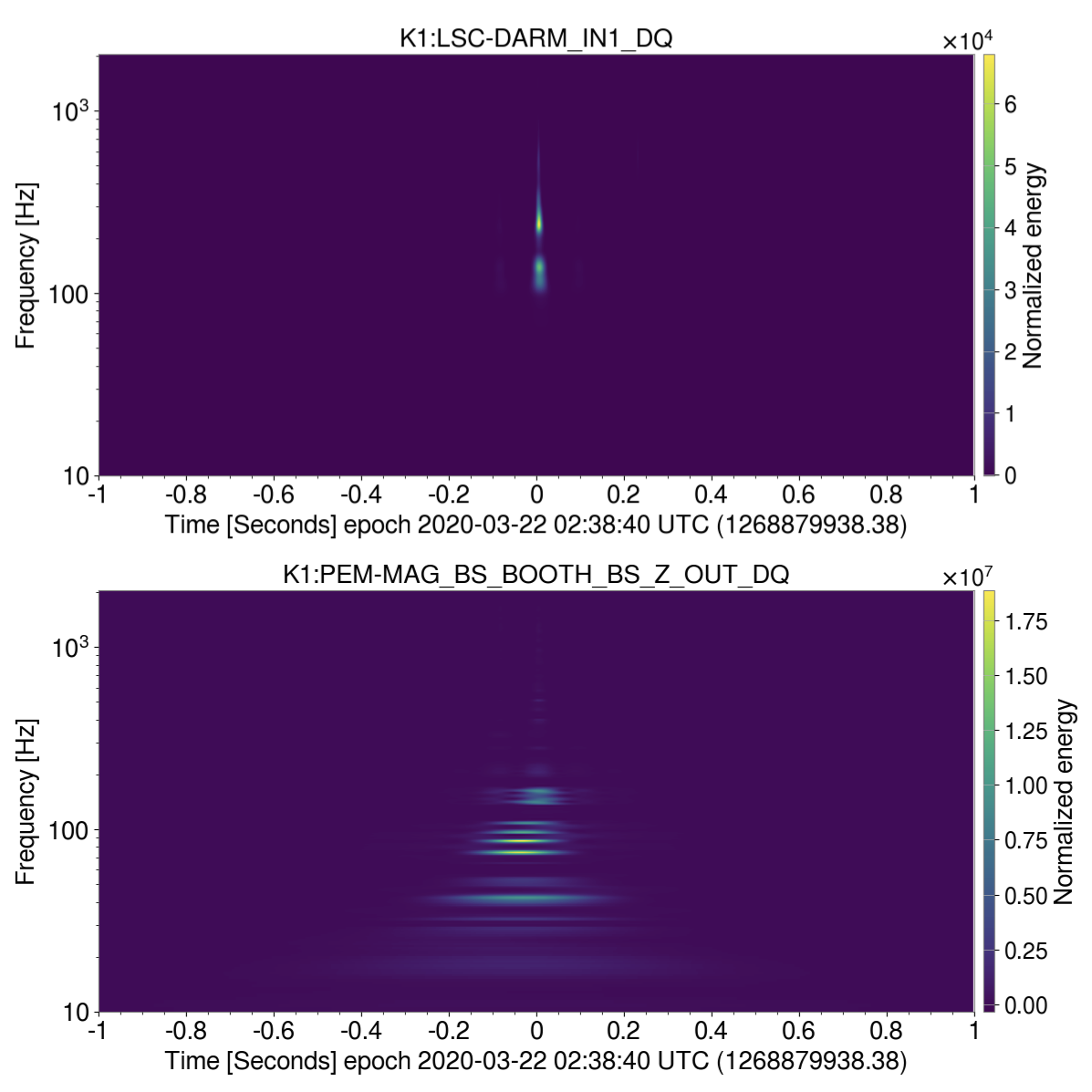}  
\caption{\small Omega scan plots of the \ac{gw} strain channel (top) and the magnetometer channel at the beam splitter (BS) station (bottom).} \label{Fig.qscan}
\end{center}
\end{figure}
First, {\it CAGMon} tool can be applied to investigate a magnetic field transient noise caused by a lightning stroke event observed by the KAGRA collaboration that was verified by a clear mutual relationship between two channels in \cite{Washimi:2021ogz}. A lightning stroke is a locally enormous release of electromagnetic waves, which clearly affects to \ac{gw} detector by a sudden variance of magnetic fields. Note that the relevant studies in LIGO, Virgo, and KAGRA are shown in terms of a short duration magnetic fields effect \cite{Kowalska-Leszczynska:2016low}, the amplified magnetic effect in the underground site \cite{Atsuta:2016wxo}, and the transient magnetic effects caused by a large scale lightning strike during GW150914 \cite{LIGOScientific:2016gtq}.

KAGRA \ac{gw} detector was influenced by a lightning stroke at 02:38:40.38 on March 21, 2020, UTC, which was recorded in the \ac{gw} strain channel and magnetometers. This detection was the first evidence observed by the KAGRA \ac{gw} detector that the lightning strokes in the atmosphere would be able to affect the underground-based GW detector within the detection range. The disturbance caused by the electromagnetic fields from the lightning stroke can affect the \ac{gw} strain channel of the KAGRA \ac{gw} detector, producing a coincident transient noise in the strain channel with a significant association. Hence, we can infer that there exists a significant correlation between the \ac{gw} strain channel and the magnetometer-related channels. The omega-scan spectrogram plots based on Q-transform are presented in Fig. \ref{Fig.qscan}.

\begin{figure*}[t!]  
\begin{center}
\includegraphics[width=\textwidth]{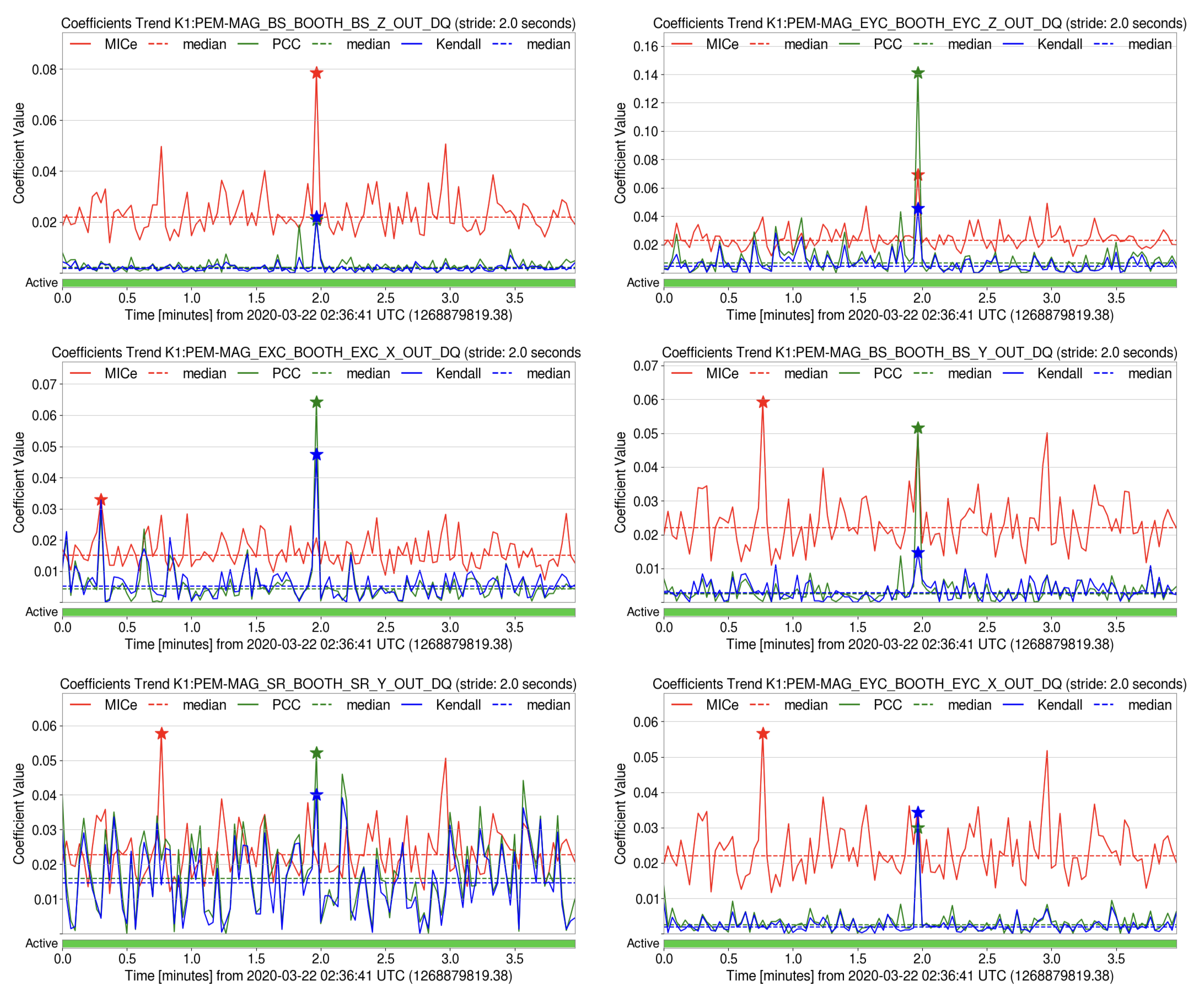}  
\caption{\small Correlation trend plot of association between \ac{gw} interferometer channel and magnetometer associated channels at BS station at the lightning event time. The solid line and the dashed line indicate the strength of the correlation coefficient and median value, respectively. Here, the star-mark represents the maximum value of $({\rm MICe}, \rho, \tau)$.} \label{Fig.lightningCAGMon}
\end{center}
\end{figure*}

\renewcommand\arraystretch{1.6}
\renewcommand{\tabcolsep}{3.8pt}
\begin{table*}[t!]
\centering
{\footnotesize
\begin{tabular}{c|l|c|c|c|c|c|c} 
\hline\hline
Event time(GPS) & Associated auxiliary channels\footnote{All exhibited channels here are related to magnetometers installed in several areas nearby the KAGRA \ac{gw} detector.} & MICe & Med(MICe)\footnote{The median value was computed for four minutes duration segment.} ($10^{-2}$) & $\rho$ & Med($\rho$)($10^{-2}$) & $\tau$ & Med($\tau$)($10^{-2}$) \\ 
\hline\hline
\multirow{6}{*}{\begin{tabular}[c]{@{}l@{}}March 22, 2020\\ 02:38:39-41UTC\\(1268879937.38\\-1268879939.38)\end{tabular}} & K1:PEM\footnote{PEM stands for the physical environment monitor that is related to subsystems for sensing environmental changes such as seismomter, magnetometer, accelerometer, and so on \cite{10.1093/ptep/ptab018}.}-MAG\_BS\_BOOTH\_BS\_Z\_OUT\_DQ & 0.079 & 2.210 & 0.021 & 0.212 & 0.022 & 0.172 \\
 & K1:PEM-MAG\_BS\_BOOTH\_BS\_Y\_OUT\_DQ & 0.050 & 2.210 & 0.052 & 0.250 & 0.015 & 0.275 \\
 & K1:PEM-MAG\_BS\_BOOTH\_BS\_X\_OUT\_DQ & 0.026 & 2.188 & 0.040 & 0.266 & 0.001 & 0.197 \\
 & K1:PEM-MAG\_EXC\_BOOTH\_EXC\_X\_OUT\_DQ & 0.021 & 1.499 & 0.064 & 0.423 & 0.047 & 0.521 \\
 & K1:PEM-MAG\_EYC\_BOOTH\_EYC\_Z\_OUT\_DQ & 0.069 & 2.309 & 0.141 & 0.709 & 0.045 & 0.474 \\
 & K1:PEM-MAG\_SR\_BOOTH\_SR\_Z\_OUT\_DQ & 0.022 & 2.271 & 0.052 & 1.595 & 0.040 & 1.458 \\
\hline
\end{tabular}
}
\caption{\small Correlation values of MICe, $\rho$, and $\tau$ at the lightning stroke event time and some associated channels with the magnetometer in the KAGRA \ac{gw} detector.}
 \label{tab:striketable}
\end{table*}

Here, we used two seconds stride of data with $8192$ data size from \ac{gw} strain channel to investigate a correlation between short-duration data segments. The mutual correlations between the \ac{gw} strain channel and the magnetometer-related channels detect a meaningful signal at the event time of the lightning stroke. Here, correlation trend plots are depicted in Fig. \ref{Fig.lightningCAGMon}, yielding a clear peak signal at the event time of the lightning stroke. It is inferred that the magnetic field noise from the lightning stroke event can affect the \ac{gw} strain channel. Interestingly, the aspect of each correlation exhibited in Fig. \ref{Fig.lightningCAGMon} shows that they are linearly or non-linearly correlated with each other for the individual subsystems. The values with significant correlations are much greater than the median values of four minutes duration data segment. In addition, we list the magnetometer-associated subsystems with a significant correlation to the lightning stroke event in Table. \ref{tab:striketable}.

\subsection{Periodic Noises from Air compressors}  
\label{sec:aircompressor}

\begin{figure}[t!]  
\begin{center}
\includegraphics[width=\columnwidth]{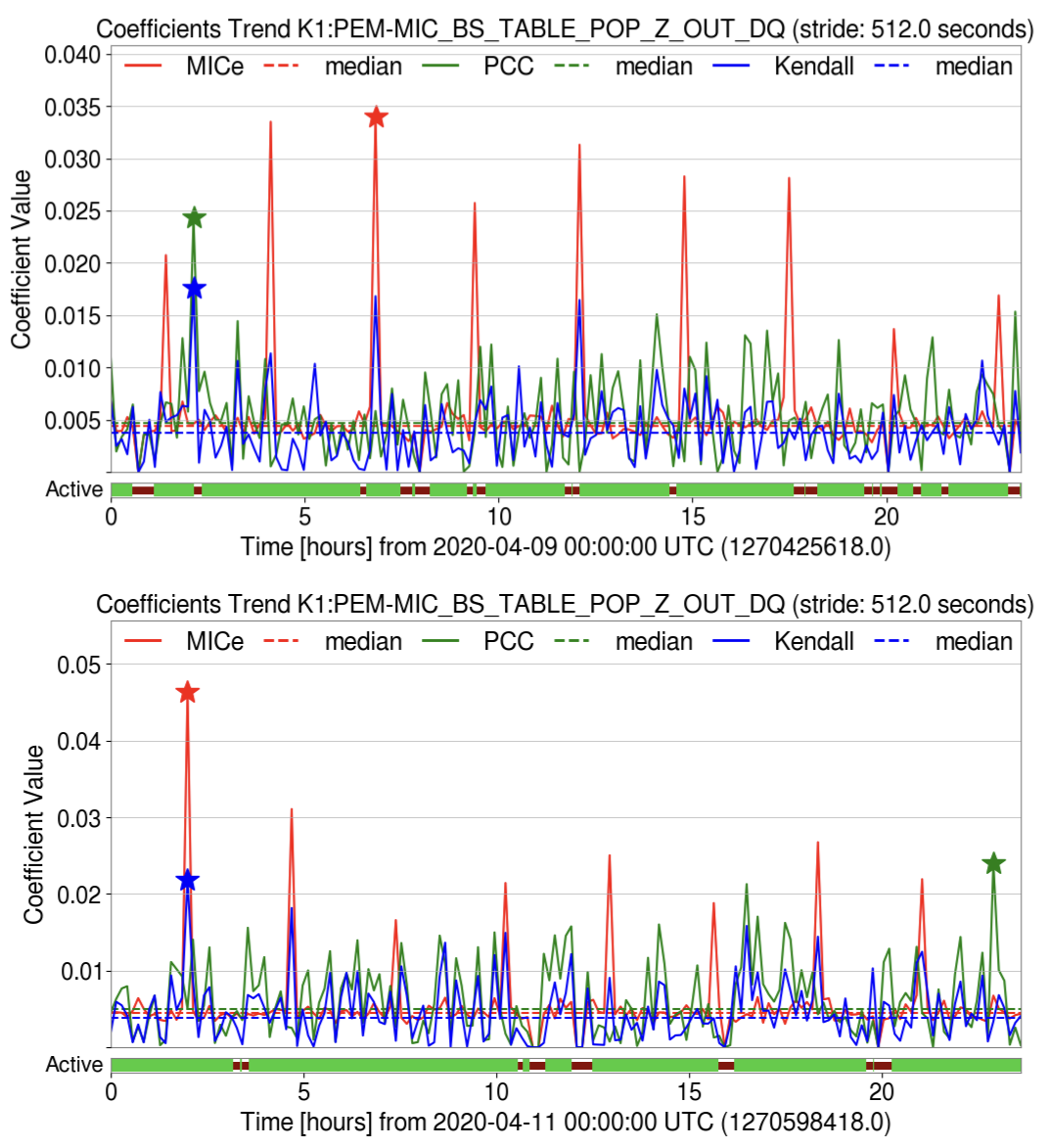}  
\caption{\small Correlation trend plots between the BNS range and microphone channels installed in the BS Station POP room on April 9, 2020 (top) and April 11, 2020 (bottom); Relatively strong MICe correlations were repeated every 2.58 hours.} \label{Fig.aircompressor}
\end{center}
\end{figure}

\begin{figure}[ht!]  
\begin{center}
\includegraphics[width=\columnwidth]{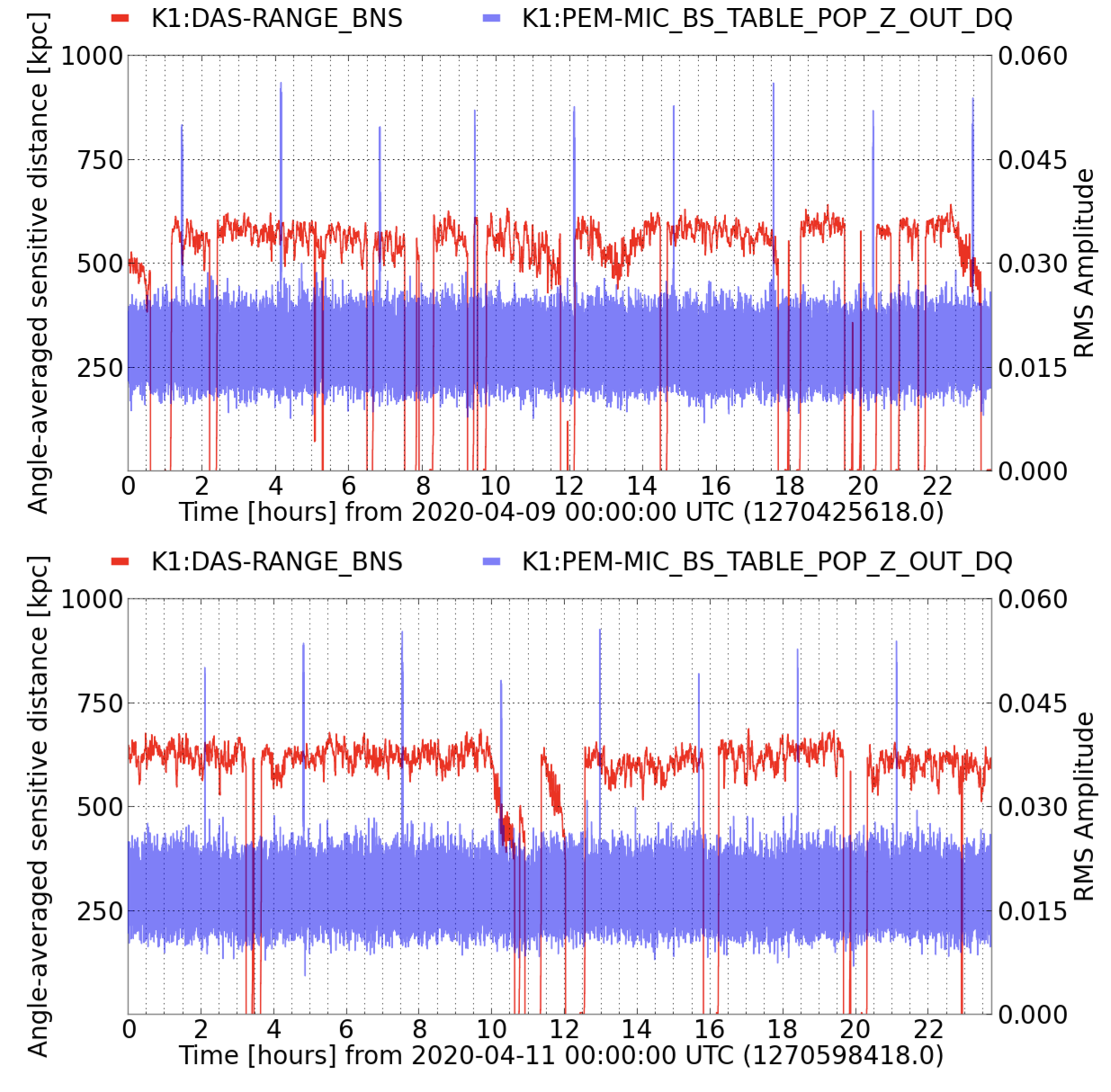}  
\caption{\small The BNS range curve experiences a sudden drop of as much as about 50kpc owing to the air compressor vibration from the OMC station on April 9, 2020 (top) and April 11, 2020 (bottom).} \label{Fig.dropAC}
\end{center}
\end{figure}

\begin{figure*}[ht!]  
\begin{center}
\includegraphics[width=\textwidth]{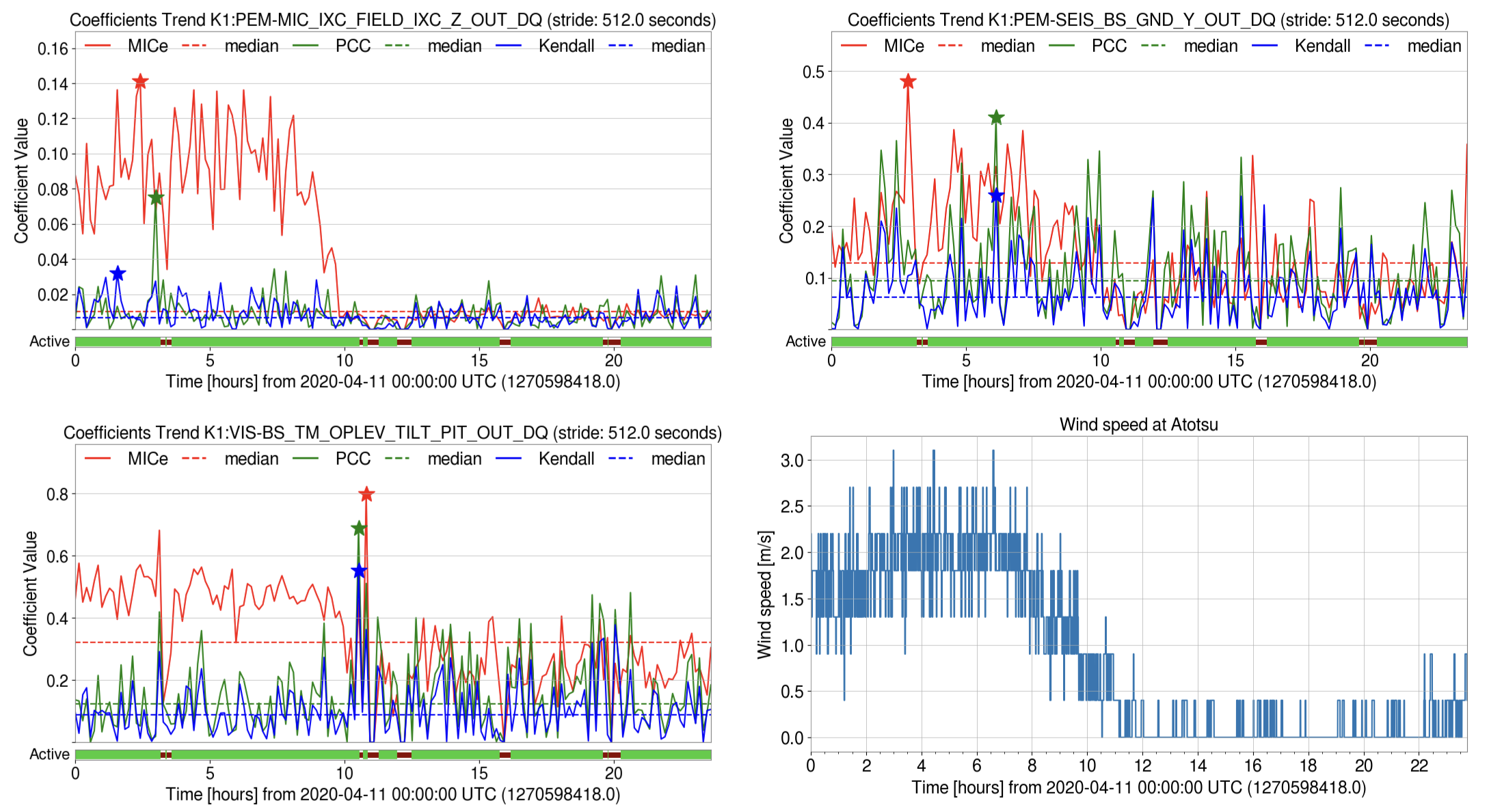}  
\caption{\small Correlation trend plots between the BNS range channel and microphone channel installed in the initial x-arm tank (top-left), the test mass optical leverage at the BS VIS channel (bottom-left), and the y-axis seismometer at BS (top-right). And, the plot of wind speed is defined as the average wind speed over two minutes at the entrance of KAGRA (Atotsu) on April 11, 2020 (bottom-right)} \label{Fig.infra}
\end{center}
\end{figure*}

We applied the {\it CAGMon} tool to the observing run of KAGRA (O3GK run) from April 7, 2020, to April 21, 2020 \cite{LIGOScientific:2022myk, KAGRA:2022fgc}. Unlike the aforementioned analysis for transient noise, we used the binary neutron star (BNS) range channel instead of the \ac{gw} strain channel to investigate the coherence between the sudden drops in the BNS range plot and the air compressor noises. The BNS range channel is a monitoring channel that represents the volume/orientation-averaged distance based on the \ac{gw} signals with a matched filtered signal-to-noise ratio (SNR) of $8$ from the two $1.4M_{\odot}$ neutron stars in a \ac{gw} detector \cite{PhysRevD.47.2198, KAGRA:2013pob}. The slowly-moving behavior caused by long-duration noise disturbances in \ac{ligo} detector has been investigated via the least absolute shrinkage and selection operator (LASSO) regression, where the BNS range channel was used \cite{Walker:2018ylg}.

In this analysis, we examined internal or external influences on the detector's sensitivity during the long period of observational mode. Because data record a minute trend of sensitivity range with around three minutes latency, we set 512 seconds of stride and 8192 of data size.
With this configuration, significant correlations were periodically repeated between the BNS range channel and microphone channels near the beam splitter (BS) station. These correlated peak signals have a harmonic frequency of 26.5Hz and repeat with a periodicity of 2.58 hours every day. Figure \ref{Fig.aircompressor} exhibits the correlation trend plots of this periodic behavior using the {\it CAGMon} tool. In addition,  it has been found that the air compressor at the output mode cleaner (OMC) produces periodic vibrations that correlated through nearby instruments to result in the sudden drop of the BNS range of the KAGRA sensitivity curve. The BNS range drop owing to the periodical air compressor noises is demonstrated in Fig. \ref{Fig.dropAC}. This phenomenon was previously reported during the KAGRA engineering operation and reduced significantly via the installation of vibration isolation and soundproof devices in the air compressors. However, the {\it CAGMon} tool caught this effect even if it had been reduced.

Consequently, this example indicates that association may exist if the trend of the coefficients is changed regardless of their strength.

\subsection{Glimpse of Acoustic Noise from Winds} 
\label{sec:infrasound}

Another interesting point via the {\it CAGMon} tool was observed by a clear correlation between the BNS range channel and the microphone/vibration isolation channels. The correlation trend plot exhibits a meaningful association from morning (9 A.M, JST) to evening (7 P.M, JST) every day in the spring season, which is inferred by a relationship with strong wind power during this time (Fig. \ref{Fig.infra}). Because the KAGRA \ac{gw} detector is installed inside the underground tunnel in Mt. Ikeno, we infer that this is responsible for the propagation of an acoustic wave noise owing to strong winds between Mt. Ikeno. Fig. \ref{Fig.infra} exhibits an interesting feature of MICe because this association only was detected by \ac{mic} rather than \ac{pcc} and \ac{kendall}; one can infer that the effect of winds can affect non-linearly the KAGRA detector. A theoretical possibility of influences by the fluctuating gravity gradient noises in the ground and air around the detector has been studied in \cite{Creighton:2000gu}. For this reason, it is worthwhile studying because the seismic coupling and up-conversion effect may affect the \ac{gw} detection band for the underground detector. 

A potential scenario of this effect is considered by a simple simulation of the elastic wave equation in the underground like the KAGRA detector. The wall of the L-shape tunnel in the KAGRA detector consists of the concrete material and air inside the tunnel. The $y$-arm of the KAGRA detector is parallel to the slop of the valley in Mt. Ikeno whereas the $x$-arm is located from the slope to deep inside the mountain. The seismic incident wave generated from the strong winds striking the slope vibrates and propagates then transforms into an acoustic wave inside the tunnel. This acoustic vibration produces sound pressure noise inside the tunnel. Because of the location of both arms in the KAGRA detector, the sound pressure noise level in the $y$-arm direction is much more severe than that in the $x$-arm direction due to the attenuation inside the deeper region of the tunnel. 

\begin{figure*}[ht!]  
\begin{center}
\includegraphics[width=\textwidth]{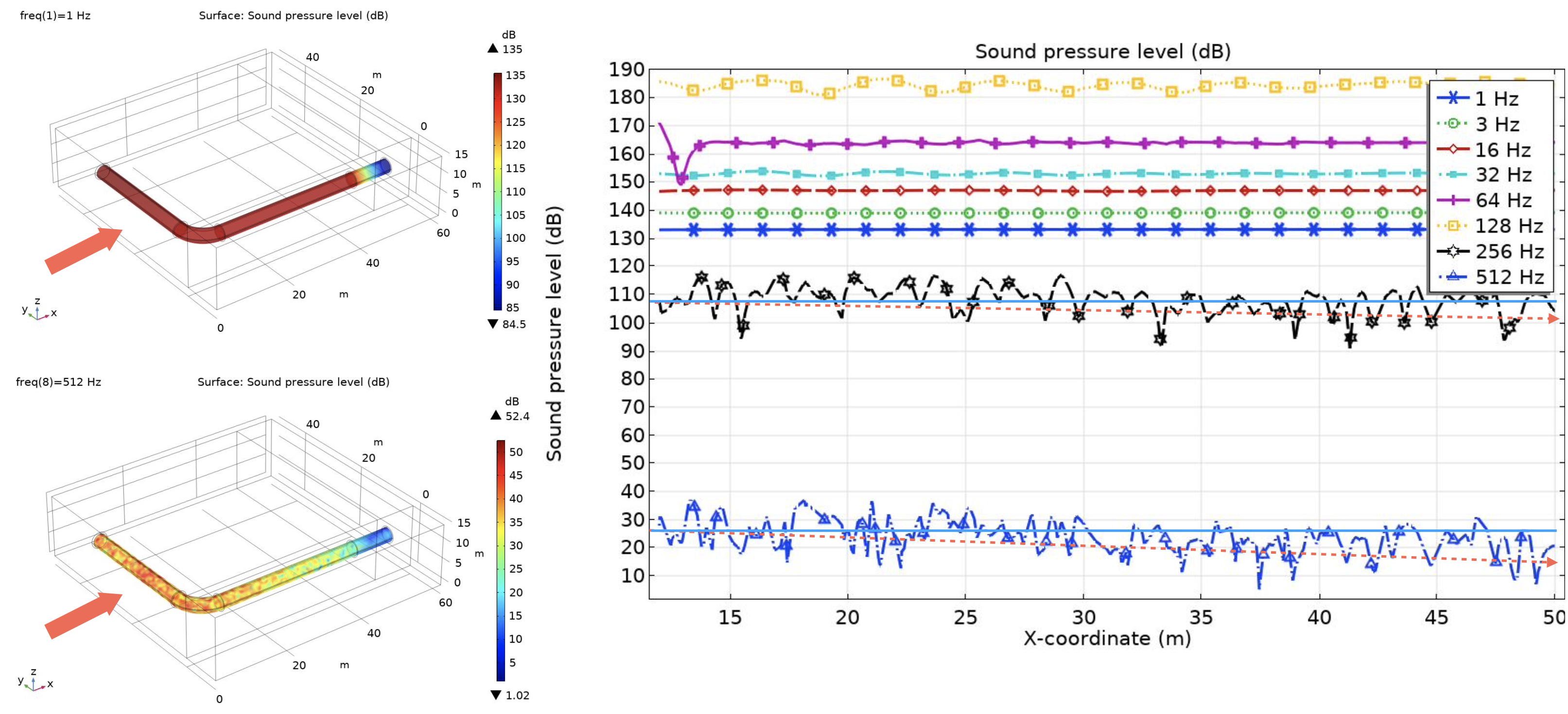}  
\caption{\small The simulation of seismic and acoustic wave propagation in the underground tunnel of KAGRA and its $x$-arm sound pressure level in diverse frequencies. The perfectly matched layer (PML) was applied in the end region of $x$-arm. } \label{Fig.ikeno}
\end{center}
\end{figure*}

The finite element method (FEM) simulation of the seismic and acoustic waves with a multi-physics configuration and the sound pressure level along $x$-arm direction in diverse frequencies are shown in Fig. \ref{Fig.ikeno}.\footnote{The simulation has been performed by COMSOL-Multiphysics 5.6 with somewhat {\it ad hoc} parameters and assumptions to check the possibility of the scenario.} 
Consequently, the seismic vibration from the slope can propagate to air fluctuation in the tunnel of the KAGRA, then producing acoustic noises inside the tunnel, which needs to be verified by more accurate simulation and instrumental measurement.

\section{Discussions}
\label{sec:discussion}
We developed a novel tool for identifying and diagnosing data association between two variables to investigate presumably correlated events between multichannels of the \ac{gw} detectors, called {\it CAGMon}. In this tool, three linear and non-linear measures, \ac{pcc}, \ac{kendall}, and \ac{mic}, were adopted and the optimized parameter selection for \ac{mic} was referred in \cite{Jung:2021jss}. 

We applied this tool for the transient and periodic noises caused by a lightning stroke event and an acoustic noise caused by an air compressor in the KAGRA \ac{gw} detector, respectively. Consequently, we verified that several magnetometer-associated subsystems influenced by the lightning stroke event have a significant association with the \ac{gw} strain channel. On the other hand, we found that the sudden detection range reductions in the BNS range curve of the KAGRA detector are associated with the periodic noise every 2.58 hours appearing in the microphone channels. The noise was identified with a harmonic frequency of 26.5Hz and the evident cause was confirmed by acoustic noises from the air compressor nearby the BS station. Finally, it is observed a glimpse of the wind effect on the underground detector via {\it CAGMon} tool.  In the daytime, the air nearby mountains become heated and makes relatively strong wind between mountains. This wind hits the ground surface where the detector locates, yielding and propagating microseismic noise and infrasound waves toward the underground detector \cite{1977JAtS}. Thus, the detector experiences acoustic and seismic vibrations during windy times. This scenario seems plausible but needs to verify in various aspects. Consequently, the seismic vibration from the slope can propagate to air fluctuation in the tunnel of the KAGRA, then producing acoustic noises inside the tunnel, which needs to be verified by more accurate simulation and instrumental measurement.

Potentially, the {\it CAGMon} tool and its application will help to overcome several limitations in the KAGRA detector, and thereby contribute not only to improving the noise reduction study but also to developing advanced tools and interfaces for the next-generation gravitational-wave telescopes. Furthermore, this tool will help scientists in the \ac{gw} detection as well as other fields of sciences. 
\acknowledgments 
The authors would like to thank the anonymous reviewer for all the invaluable comments and suggestions, which helped us improve the manuscript's quality. The authors also would like to thank Kyujin Kwak, Kyungmin Kim, Sangwook Bae, Hyungwon Lee, and Whansun Kim for their helpful discussions and comments.
This work was supported by the Basic Science Research Program through a National Research Foundation of Korea (NRF) funded by the Ministry of Education (NRF-2020R1I1A2054376). Furthermore, this work is partially supported by the NRF grant funded by the Korean government's Ministry of Science and ICT (No.\ 2019R1A2C2006787, No.\ 2016R1A5A1013277,  No.\ 2019R1C1C1010571, and No.\ 2021R1A2C1093059). 

\bibliographystyle{apsrev4-1}
\bibliography{Ref}



\end{document}